\documentclass[12pt]{article}
\usepackage{latexsym}
\begin{document}
\input epsf
\def\be{\begin{equation}}
\def\bea{\begin{eqnarray}}
\def\ee{\end{equation}}
\def\eea{\end{eqnarray}}
\def\d{\partial}
\def\la{\lambda}
\def\eps{\epsilon}
\def\a{{\cal A}}
\def\q{\quad}
\def\b{\bigskip}
\begin{flushright}
OHSTPY-HEP-T-02-007\\
hep-th/0206107
\end{flushright}
\vspace{20mm}
\begin{center}
{\LARGE  Rotating deformations of $AdS_3\times S^3$, the orbifold CFT and
strings in the pp-wave limit}
\\
\vspace{20mm}
{\bf  Oleg Lunin  and  Samir D. Mathur \\}
\vspace{4mm}
Department of Physics,\\ The Ohio State University,\\ Columbus, OH 43210, USA\\
\vspace{4mm}
\end{center}
\vspace{10mm}
\begin{abstract}

We construct an exact metric which  at short distances is the metric
of massless particles in 5+1
spacetime (moving along a diameter of the sphere) and is $AdS_3\times
S^3$ at infinity.
We also consider  a set of a conical defect spacetimes which are
locally $AdS_3\times S^3$ and have the
masses and charges of a special set of chiral  primaries of the dual
orbifold CFT.   We find that excitation energies for a scalar field 
in the latter geometries agree exactly
with the excitations in the corresponding CFT state created by  twist 
operators: redshift in
the geometry reproduces `long circle' physics in the CFT.    We 
propose a map of  string states in
$AdS_3\times S^3\times T^4$ to states in the orbifold CFT, analogous
to the recently discovered map for
$AdS_5\times S^5$. The vibrations of the string can be pictured as
oscillations of a Fermi sea in the
  CFT.

\end{abstract}
\newpage

\section{Introduction.}
\renewcommand{\theequation}{1.\arabic{equation}}
\setcounter{equation}{0}

String theory on $AdS_5\times S^5$ is believed to be dual to $N=4$
Super-Yang-Mills gauge theory on
the boundary of
$AdS_5$  \cite{AdSCFT}. Weakly coupled gauge theory is dual
to a string background that is not
well described by perturbative supergravity. Nevertheless it is
possible to demonstrate  several
relations between the two dual theories.

String theory on $AdS_3\times S^3\times M_4$ is expected to be dual
to a  2-dimensional CFT on the
boundary of $AdS_3$. It has been conjectured that this CFT can be
represented as a deformation of a
2-dimensional sigma model with the orbifold target space $M_4^N/S_N$
\cite{orbCFT}. (Here $S_N$ is the
permutation group for $N$ variables.)  The analogue of free
Yang-Mills would be the `orbifold point'
where we have the orbifold with no deformation. This orbifold model
has given several exact
agreements with the dual theory  for both BPS and near-BPS quantities.

An essential feature of the orbifold theory is the existence of
`twist operators' $\sigma_n$. The CFT is
described by $N$ copies of a $c=6$ CFT. The twist $\sigma_n$ links
together $n$ of these copies to yield a
$c=6$ CFT living on a circle that is $n$ times longer than the circle
on which the initial CFT was defined.
The excitations of the CFT on this long circle can have very low
energy if $n$ is large.

In this paper we study two related questions that arise in this
duality map. A chiral primary in the CFT
is described by a twist $\sigma_n$ which also carries a  charge that
makes $h=j$. We call the resulting chiral primary $\sigma_n^{--}$.
Consider the state
where all the $N$ copies of the CFT are twisted so as to make the
state $(\sigma^{--}_n)^{N/n}|0\rangle_{NS}$.
Since all copies of the CFT now live on `long circles' we must have
low energy excitations of the CFT, and
we must find that in the dual gravity theory the supergravity fields must have
corresponding low energy eigenmodes.

For the CFT in the Ramond (R) sector this issue was studied in
\cite{lm4}, where supergravity duals
were constructed for all the R ground states -- these states are in
one-to-one correspondence with the
chiral primaries of the NS sector which in turn are  of the form
$\sigma^{--}_{n_1}\sigma^{--}_{n_2}\dots \sigma^{--}_{n_k}$. It was found
that the
length of the CFT circles was
described in the supergravity dual by the {\it depth} of the `throat'
of the near horizon geometry.
The simplest geometries were dual to configurations of the type
$(\sigma_n^{--})^{N/n}$, and for these
geometries the travel time around the CFT `long circle' agreed {\it
exactly} with the travel time down the
throat of the gravity solution.

In our present study of the NS sector we therefore look for the
gravity  duals of the CFT states
$(\sigma_n^{--})^{N/n}|0\rangle_{NS}$. We solve the wave-equation in the
geometries, and find energy
levels that agree exactly with the  energy levels for the CFT with
`long circles'.

The other  question relates to the recent proposal by Berenstein,
Maldacena and Nastase
\cite{bmn} that strings moving fast on the $S^5$ in $AdS_5\times S^5$
can be described easily in the
dual gauge theory.  A supergravity quantum moving fast on the $S^3$
in $AdS_3\times S^3\times
M_4$ is described by a chiral primary $\sigma^{--}_n$  (or its SUSY
descendent) with a large value of $n$.  We
note that
$\sigma^{--}_n$ can be written in the form $(\sigma^{--}_2)^n$, and use the
`bits' $\sigma^{--}_2, J^-_0\sigma^{--}_2, \bar J^-_0\sigma^{--}_2$
as analogs of the
$Z,X_i$ of \cite{bmn} to construct the low lying string excitations
of the quantum.

In more detail we do the following:
\b

(a)\q The chiral primary $\sigma_n^{--}$ in the CFT is described by a
supergravity
quantum having angular momentum $n$ on the $S^3$. Thus the state
$(\sigma_n^{--})^{N/n}|0\rangle_{NS}$
should be described by a collection of such quanta. If $n\gg 1$ the
quanta will be confined to a narrow
width around the circle of rotation, and if we also have $N/n\gg 1$
then the number of quanta will be
large and  the resulting geometry should be well described by a
supergravity solution.

In flat space the metric produced by a massless particle is the
Aichelburg--Sexl metric \cite{AS70}, and we can
readily extend this to describe a  set of particles  uniformly
distributed along the line of motion. We
construct an exact solution that goes over to the analogue of the
Aichelburg--Sexl solution in 5+1
spacetime near the moving particles, and to $AdS_3\times S^3$ at
infinity. Just as in the Aichelburg--Sexl
case, the solution at the linearized level turns out to be exact.

\b
(b)\q We do not however find any obvious way to separate the scalar
wave equation in the above
metric,  in contrast to the separability found in the R sector
solutions \cite{larsen, lmTube}. Further the
travel time for quanta in this geometry from infinity to the center
does not agree with the expected
time for travel around the `long circles' of the CFT state
$(\sigma_n^{--})^{N/n}|0\rangle_{NS}$. We thus look
for a different metric that could be dual to these chiral primaries.
We consider a set of metrics with
a conical defect singularity along a circle; such  conical defect
singularities arose in the R sector solutions
and were studied in \cite{bal, mm, mathur}. For this class of metrics
we find that the scalar wave
equation separates, and the energy levels for the $l=0$ harmonic of
the scalar match {\it exactly} the
corresponding energy levels in the dual CFT on the `long circle'.
The travel time for a quantum from
infinity to the center and back also agrees exactly with the travel
time around the long circle of the CFT.

These results indicate that for the massless quanta describing the
chiral primary
$(\sigma_n^{--})^{N/n}|0\rangle_{NS}$ the lowest energy solution is not
the one with an Aichelburg--Sexl
type singularity but rather the one with the conical defect singularity.

\b
(c)\q We next consider a single quantum with high angular momentum
in the geometry
$AdS_3\times S^3\times M_4$. The quantum is described by a  chiral
primary $\sigma_n^{--}$, but the
remaining $N-n$ copies of the $c=6$ CFT are not twisted into `long
circles'. The chiral primary $\sigma_n^{--}$
can be written as $(\sigma^{--}_2)^N$, where $\sigma^{--}_2$ is a chiral
primary that can be regarded as a
building black to make all higher chiral primaries. We argue that
$\sigma^{--}_2$ plays the role of the field $Z$
in \cite{bmn}, and that $J_0^-\sigma^{--}_2, \bar J_0\sigma^{--}_2$ are the
analogs of the variables $X_i$ that
          describe stringy excitations around the supergravity quantum.  We
find that the CFT description of
these string excitations can be interpreted as low energy vibrations
of a `Fermi fluid' which arises from
the fermionic excitations that give $\sigma_n^{--}$ its charge.

\bigskip

While we were working on this paper there appeared the paper \cite{japanese}
which overlaps partly with our discussion in
section 5.

\section{The Aichelburg--Sexl type solution in $AdS_3\times S^3$}
\renewcommand{\theequation}{2.\arabic{equation}}
\setcounter{equation}{0}

We begin by recalling the metric produced by a massless point
particle moving along the $z$ direction in $3+1$ dimensions \cite{AS70}.
\be
ds^2=-dt^2+dx^2+dy^2+dz^2+8p\delta(t-z)\log(x^2+y^2)
(dt-dz)^2
\ee
If we consider instead a set of such particles distributed
uniformly along the $z$-axis then we get a time-independent metric:
\be
ds^2=-dt^2+dx^2+dy^2+dz^2+q\log(x^2+y^2)(dt-dz)^2.
\ee
Note that the function $\log (x^2+y^2)$ appearing in the
metric is a harmonic function
of the transverse coordinates $x,y$  with $\delta$-function source at
the singular line
$x=y=0$. In 5+1 dimensions the corresponding  metric will be
($x_i=x_1\dots x_4$)
\be\label{5DflatAS}
ds^2=-dt^2+dz^2+dx_idx_i+\frac{q}{(x_i x_i)}(dt-dz)^2,
\ee

We wish to construct solutions that describe a set of massless
particles rotating along a
diameter  of $S^3$ in the space $AdS_3\times S^3$.  This metric
arises in IIB string theory
after compactification on a 4-manifold $M_4$ which can be $T^4$ or
$K3$.  We will take the case
of $T^4$ for concreteness in this paper, though there is no
fundamental difference if we were
to take $K3$ instead.  The curvature is produced by the 2-form gauge
field $B^{NSNS}_{\mu\nu}$;
we will just call this field $B_{\mu\nu}$ hereafter.
For unperturbed $AdS_3\times S^3 \times T^4$ the 10-d Einstein metric
is\footnote{The action for the relevant fields is
$\int \sqrt{-g}~[R-{1\over 12} H^2]$ and
$H_{\mu\nu\lambda}=\partial_\mu B_{\nu\lambda}
+\partial_\nu B_{\lambda \mu}+\partial_\lambda B_{\mu\nu}$.}
\be
ds^2=-(1+\frac{r^2}{L^2})dt^2+\frac{dr^2}{1+\frac{r^2}{L^2}}+
r^2 d\chi^2
+L^2\left\{d\theta^2+\cos^2\theta d\psi^2+\sin^2\theta d\phi^2\right\}
+\sum_{i=1}^4 dz_i dz_i
\ee
and the gauge field is\footnote{Thus $B_{\phi\psi}=L^2\cos^2\theta$ etc.}
\be
B=L^2\cos^2\theta d\phi\wedge d\psi+\frac{r^2}{L}dt\wedge d\chi.
\ee

We consider a set of massless particles rotating along the diameter
$\theta=0$ of the $S^3$
at the location $r=0$ in the $AdS_3$. (We will ignore the $T^4$ in
what follows -- one may
assume that the particle wavefunctions are uniformly smeared along the
$T^4$.) We thus look
for a solution of the IIB field equations that behave as (\ref{5DflatAS})
near  $\theta=0, r=0$, and go
over to $AdS_3\times S^3$ at large $r$.

We first construct the linear solution corresponding to  small
strength of the source at
$\theta=0, r=0$. It turns out that just as is the case for the
Aichelburg--Sexl metric in 3+1
dimensions, the linear solution is exact. The full 10-d metric is
\bea\label{NewASmetr}
d s^2&=&-(1+\frac{r^2}{L^2})dt^2+\frac{dr^2}{1+\frac{r^2}{L^2}}+
r^2 d\chi^2
+L^2\left\{d\theta^2+\cos^2\theta d\psi^2+\sin^2\theta d\phi^2\right\}
+\sum_{i=1}^4 dz_i dz_i\nonumber\\
&+&\frac{q}{r^2+L^2\sin^2\theta}\left[\left\{(1+\frac{r^2}{L^2})dt+
L\cos^2\theta d\psi\right\}^2-L^2
\left\{\frac{r^2}{L^2}d\chi-\sin^2\theta d\phi\right\}^2\right]
\eea
and the NS--NS two--form field is
\bea\label{NewASfield}
B&=&L^2\cos^2\theta d\phi\wedge d\psi+\frac{r^2}{L}dt\wedge d\chi\nonumber\\
&-&\frac{q}{L}\frac{1}{r^2+L^2\sin^2\theta}
\left\{(1+\frac{r^2}{L^2})dt+
L\cos^2\theta d\psi\right\}\wedge\left\{
r^2d\chi-L^2\sin^2\theta d\phi\right\}
\eea

\section{Metrics with conical defects}
\label{SectCone}
\renewcommand{\theequation}{3.\arabic{equation}}
\setcounter{equation}{0}

In \cite{bal}, \cite{mm} a set of metrics was considered which could
be obtained as special
cases of metrics for rotating charged holes found in \cite{cy}. The
metrics describe the D1-D5
bound state carrying angular momentum, and go over to  flat space at
spatial infinity. In the near horizon region (the `$AdS$ region'),
       the solutions have the form
\bea
\label{two}
ds^2&=&-(\frac{r^2}{L^2}+\gamma^2)dt^2+r^2d\chi^2+
\frac{dr^2}{\frac{r^2}{L^2}+\gamma^2}\nonumber\\
&+&L^2\left\{d\theta^2+\cos^2\theta (d\psi-\gamma d\chi)^2+
\sin^2\theta (d\phi-\gamma \frac{dt}{L})^2\right\},\\
B&=&\frac{r^2}{L}dt\wedge d\chi+
\cos^2\theta (d\psi-\gamma d\chi)\wedge (d\phi-\gamma\frac{dt}{L})
\eea
It was argued in \cite{bal, mm} that these solutions describe ground
states of the D1-D5 system
in the Ramond sector.  It was found in \cite{lm4} that only a special
class of Ramond ground
states are described by these solutions. The solution for the
general ground state was
then constructed, and it was found that the singularity had the shape
of a complicated curve
for a generic metric.

On the other hand we can write down a larger class of metrics
similar to the simple form
(\ref{two})\footnote{Such a class of metrics was studied in \cite{mathur}}:
\bea
\label{three}
ds^2&=&-(\frac{r^2}{L^2}+\gamma^2)dt^2+r^2d\chi^2+
\frac{dr^2}{\frac{r^2}{L^2}+\gamma^2}\nonumber\\
&+&L^2\left\{d\theta^2+\cos^2\theta (d\psi-{\tilde\beta}
\frac{dt}{L}-\alpha d\chi)^2+
\sin^2\theta (d\phi-{\tilde\alpha}\frac{dt}{L}-\beta d\chi)^2\right\},\\
B&=&\frac{r^2}{L}dt\wedge d\chi+
\cos^2\theta (d\psi-{\tilde\beta}\frac{dt}{L}-\alpha d\chi)\wedge
(d\phi-{\tilde\alpha}\frac{dt}{L}-\beta d\chi)
\eea
All these metrics possess a conical
defect at $\theta=0, r=0$, and are locally $AdS_3\times S^3$ elsewhere.
Thus they are quite different from the metric found in the
previous section.

Due to the presence of the conical  singularity we do not know a
priori that all these metrics are solutions of
Type IIB string theory.  We will however try to identify a subclass
of such metrics with the dual of chiral
primaries of the NS sector, by studying the values of conserved
charges and the spectrum of low
energy excitations around these solutions.

      Let
us briefly discuss the charges associated with metric (\ref{three}).
The 2+1 gravity theory obtained by
reduction of IIB on
$S^3\times M_4$ has a gauge field
$A^\alpha_\mu$, where
$\alpha$ is an index on $S^3$ and $\mu$ is an index in the $AdS_3$. The field
equation for $A$ is
\be
d*dA+dA=*j
\ee
The conserved charge is thus given by the line integral over the
circle $t=t_0, r=r_0$ of the quantity $*dA+A$. The first part is the
usual flux of electrodynamics; it  vanishes at infinity for the
metrics we consider. The second part is a Wilson line, and gives the
angular momentum of the solution.
The angular momenta
corresponding to translations in
$\phi$ and
$\psi$ directions are:
\be
j_{(\psi)}=\frac{\beta L}{4G_3},\qquad j_{(\phi)}=\frac{\alpha L}{4G_3},
\ee
where $G_3$ is three dimensional Newton's constant.
The mass of the solution is
\be
M=-\frac{\gamma^2}{8G_3}
\ee

The quantities in the gravity theory can be related to quantities in
the boundary CFT by the following relations \cite{henn,bal,mm}
\be\label{BalRelSt}
c=\frac{3L}{2G_3}.
\ee
\be\label{J0BetaGen}
j_0=\frac{1}{2}(j_{(\phi)}-j_{(\psi)}),\qquad
{\bar j}_0=-\frac{1}{2}(j_{(\phi)}+j_{(\psi)})
\ee
Let $l_0+ \bar l_0$ give the energy and $l_0-\bar l_0$ give the
$AdS_3$ momentum of a configuration.
Then to map to CFT levels is achieved by addition of a Sugawara term
and a constant shift:
\bea
L_0&\equiv& l_0+\frac{c}{24}+\frac{6(j_0)^2}{c}\nonumber\\
\label{BalRelFin}
{\bar L}_0&\equiv& {\bar l}_0+\frac{c}{24}+\frac{6({\bar j}_0)^2}{c}
\eea
For the solutions (\ref{three})
\be\label{J0Beta}
l_0={\bar l}_0=\frac{LM}{2}=-\frac{L\gamma^2}{16G_3},\qquad
j_0=\frac{c}{12}(\alpha-\beta),\qquad
{\bar j}_0=-\frac{c}{12}(\alpha+\beta)
\ee
and
\bea
L_0&=&\frac{c}{24}
(-\gamma^2+1+(\alpha-\beta)^2)\\
{\bar L}_0&=&
\frac{c}{24}(-\gamma^2+1+(\alpha+\beta)^2)
\eea

The Ramond vacuum of the CFT is dual to the geometry (\ref{two})
with\footnote{The values of
   $\tilde\alpha, \tilde\beta$ are determined by the spectral flow
parameters of the
CFT \cite{bal, mm}.}
\be
\alpha=\gamma,\quad {\tilde\alpha}=\gamma,\quad \beta={\tilde\beta}=0
\ee

{} From the charge and dimensions of the above solutions one infers that
the NS vacuum is given by
\be
\alpha={\tilde\alpha}=0, ~~\beta={\tilde\beta}=0, ~~\gamma=1
\ee
Chiral primaries in the NS sector have $L_0=j_0, \bar L_0=\bar J_0$
which corresponds to
\be
\alpha={\tilde\alpha}=0, ~~\beta={\tilde\beta}=\gamma-1.
\ee
For these chiral primaries the geometry (\ref{three}) becomes:
\bea\label{NSsolut1}
ds^2&=&-(\frac{r^2}{L^2}+\gamma^2)dt^2+r^2d\chi^2+
\frac{dr^2}{\frac{r^2}{L^2}+\gamma^2}\nonumber\\
&+&L^2\left\{d\theta^2+\cos^2\theta (d\psi-\beta\frac{dt}{L})^2+
\sin^2\theta (d\phi-\beta d\chi)^2\right\},\\
\label{NSsolut2}
B&=&\frac{r^2}{L}dt\wedge d\chi+
\cos^2\theta (d\psi-\beta\frac{dt}{L})\wedge
(d\phi-\beta d\chi),\qquad \gamma=\beta+1
\eea
For these solutions the parameter $\gamma$ is related to $j_0$ in the
following way
by eqn.  (\ref{J0Beta}).
\be\label{J0Gamma}
\gamma=1-\frac{12}{c}j_0=1-\frac{2j_0}{N}
\ee

Let us see which chiral primaries in the CFT would be dual to these
geometries. We review the structure of
chiral primaries of the $N=4$ orbifold CFT in Appendix \ref{AppRev}. Consider
the chiral primary $\sigma_n^{--}$.
This has $h=j={n-1\over 2}$. The states we are interested in are
generated by a set of such chiral primaries:
\be\label{BcgrChPr}
[\sigma_n^{--}]^{N/n}:\qquad L_0=j_0=\bar L_0=\bar
j_0=\frac{N}{n}\frac{n-1}{2}\qquad
\ee
We then find from (\ref{J0Gamma}) the value of the parameter $\gamma$
in the metrics (\ref{NSsolut1}) dual
to the state
$[\sigma_n^{--}]^{N/n}|0\rangle_{NS}$~:
\be
\gamma=\frac{1}{n}
\ee

\section{Energy levels and travel time for scalar fields.}
\label{spectrum}
\renewcommand{\theequation}{4.\arabic{equation}}
\setcounter{equation}{0}

The state in the CFT created by the chiral primary
$\sigma_n^{--}|0\rangle_{NS}$  is described in the dual theory by a
supergravity particle
in
$AdS_3\times S^3\times M_4$
\cite{exclusion}. Thus the state $(\sigma_n^{--})^{N/n}|0\rangle_{NS}$
should be described  by a collection of  $N/n$ such particles. If $n\gg 1$ the
particles have a high angular momentum, and their  wavefunctions
will be confined to a very narrow width around
the diameter on the
$S^3$ on which they rotate. Thus they would be pointlike particles
for supergravity. If we also have $N/n\gg 1$ then we expect to
describe the resulting physics by a classical supergravity solution.

The  configuration constructed in the above way will be naturally
smeared uniformly along the diameter of rotation. We assume that
all wavefunctions are constant over the $M_4$, and so can restrict
attention to $AdS_3\times S^3$. We thus have a line of massless
particles in 5+1 dimensional spacetime.

A first guess would therefore be that near to this line of particles
the solution behaves like (\ref{5DflatAS}), the Aichelburg--Sexl solution in
flat
5+1 spacetime. We  therefore consider the exact solution
(\ref{NewASmetr}).\footnote{Note that the $B$ field (\ref{NewASfield}) has a
divergent quartic invariant
$H_{\mu\nu\la}H^{\nu\la\sigma}H_{\sigma\tau\rho}H^{\tau\rho\mu}$ near the
singular line,  though the divergence of the
invariant is much weaker than would be naively expected from the
magnitude of the
$H$ field near the singularity. }

At this point  we are led to compare this solution with the metrics
found in \cite{bal, mm, lm3} for the states of the CFT in the R
sector.
We expect to have somewhat similar properties for the R and NS cases,
since one is just a spectral flow  of the other.
In the R sector the wave equation for a scalar separated between the
angular and radial variables.  Such a separation  is not obvious
for the geometry  (\ref{NewASmetr}). Further,  in \cite{lm4} the travel time
around the `long circles' in the CFT agreed exactly
with the travel time in the throat of the dual supergravity solution.
We do not find such an agreement for the solution (\ref{NewASmetr}).   We
are therefore led to consider other possible solutions that could
describe the state $(\sigma_n)^{N/n}|0\rangle_{NS}$.

The solutions (\ref{NSsolut1}), (\ref{NSsolut2})  have conical defects that
are similar to those
found for the R sector metrics in \cite{bal, mm}. If we assume the
relations (\ref{BalRelSt})--(\ref{BalRelFin}) postulated in \cite{henn,bal}
between
quantities in the gravity theory and quantities in the CFT, then we
find
that
$L_0=j_0, {\bar L}_0={\bar j}_0$ for these solutions, so they have the
right structure to be identified as the states of the string theory
dual to the states $(\sigma_n)^{N/n}|0\rangle_{NS}$ created by chiral
primaries in the CFT.

We will now study the energy levels  of some supergravity excitations around
the metrics  (\ref{NSsolut1}) and see that they agree with the
excitation levels expected from the dual CFT.

\subsection{Energy levels for $l=0$ scalar excitations}

There are several scalars in the 5+1 dimensional theory obtained by
reduction on $M_4$.  For concreteness we take $M_4=T^4$. Let
the $T^4$ be described by coordinates $x_1\dots x_4$. Then the
graviton $h_{12}$ is a minimally coupled  scalar in the remaining
5+1 dimensions
\be
\Phi\equiv h_{12}, ~~~
\Box\Phi=0
\ee

In  Appendix \ref{AppWave} we solve this wave equation in the metric
(\ref{NSsolut1}).
The wave equation is separable, and we have solutions of the form
\be
\Phi(t,r,\chi,\theta,\psi,\phi)=\exp(-i\omega t+ip\psi+iq\phi+i\la\chi)
H(r)\Theta(\theta)
\ee

We get normalizable solutions for the frequencies (\ref{ResulFreq})
\be\label{OmegaK}
\omega_k=\frac{\beta p}{L}\pm
\left\{\frac{2\gamma}{L}(k+1+\frac{l}{2})+
\left|\frac{\la+\beta q}{L}\right|\right\}\qquad k=0,1,2,\dots.
\ee

Consider the case $l=0$, $\la=0$. The spectrum is
\be\label{OmKnoRot}
\omega_k=\frac{2}{nL}(k+1), ~~k=0,1,2,\dots,
\ee
where we have used the relation $\gamma=1/n$ for chiral primaries
(see (\ref{BcgrChPr})).

Let us relate the energy of such quanta with the expected changes in
quantities in the CFT.
For $\delta j_0=\delta \bar j_0=0$ the relations (\ref{BalRelFin}) give
\be
\delta(L_0+ \bar L_0)=\delta(l_0+\bar l_0)=L\omega_k=\frac{2(k+1)}{n}
\label{SUGRAFreq}
\ee

Now let us consider the CFT itself. The graviton $h_{12}$ is known to
be described by
a pair of excitations \cite{dasmathurtwo}
\be
h_{12}\rightarrow {1\over \sqrt{2}}[\partial X^1\bar\partial X^2+
\partial X^2\bar\partial X^1]
\ee
There is one left mover and one right mover on the `long circle' of
length $2\pi n$.  We can have fractional levels for $L_0, \bar
L_0$ but $L_0-\bar L_0$ must be an integer \cite{dasmathur,
maldasuss}. The energy levels for $L_0-\bar L_0=0$ are given by
\be
\delta L_0=\delta \bar L_0={k'\over n}, ~~ \delta(L_0+\bar
L_0)=\frac{2k'}{n}, ~~k'=1,2,\dots
\ee
so that we get exact agreement with (\ref{SUGRAFreq}).

\subsection{Travel time}

In \cite{lm4} it was found that the time $\Delta t_{CFT}$ for
excitations to travel around the `long circle' of the
CFT was exactly equal to the time $\Delta t_{SUGRA}$ taken for a
supergravity quantum to travel once up and
down the `throat' of the dual geometry. In the present case the
geometry at infinity is $AdS$ rather than flat
space.

We  observe from (\ref{OmegaK}) and the fact that $\gamma={1\over
n}$,  $\beta={1\over n}-1$ that the
frequencies $\omega_k$ have the form
\be
\omega={1\over L}[{m_1\over n}+m_2], ~~~m_1, m_2 ~{\rm integral}
\ee
Consider any wavefunction for the scalar particle made by an
arbitrary superposition of such frequencies.
Then the wavefunction returns to its initial form after a time
\be
\Delta t_{SUGRA}=n2\pi L
\label{ten}
\ee

Now consider the dual CFT.  The left and right movers are massless
excitations, and travel at the speed of light
around the spatial circle of the cylinder on which the CFT lives. This gives
$L\delta\chi=\delta t$ for the right movers and $L\delta\chi=-\delta
t$ for the left movers. But the twist $\sigma_n$ makes the circle
on which the CFT excitations live a `long circle' of length $2\pi n$
times the initial circle size of the CFT. The
state $(\sigma_n^{--})^{N/n}|0\rangle_{NS}$  that we have chosen  has
all copies of the CFT arranged into long
circles of this length. Thus any excitation of this CFT state will
return to its  initial form after a time
\be
\Delta t_{CFT}=2\pi n L
\ee
which is in exact agreement with (\ref{ten}).

To summarize, we have seen that the CFT state
$(\sigma_n^{--})^{N/n}|0\rangle_{NS}$ has all copies of the
$c=6$ CFT joined into `long circles' that  wind $n$ times  the
spatial direction before closing. The low energy
excitations in the CFT are thus of energy $\sim 1/n$. In the dual
supergravity solution we do not see any
`multiwinding' around the $\chi$ direction. What we have instead is a
deep `throat' like structure  near
$r=0$ that  leads to a large redshift between $r=\infty$ and
$r\approx 0$. Particle wavefunctions trapped
near $r=0$ have energies that are low ($\sim 1/n$) due to this
redshift. This relation  `multiwinding in CFT
$\rightarrow$ redshift in gravity' is expected to have general
validity and was found also in the R sector
computations of \cite{lm4}.

\section{CFT states representing a single string}
\renewcommand{\theequation}{5.\arabic{equation}}
\setcounter{equation}{0}

\subsection{String states in $AdS_5\times S^5$}

Let us review the idea proposed in \cite{bmn}. We consider a
supergravity particle rotating on the diameter of
$S^5$ in
$AdS_5\times S^5$, at the origin of $AdS_5$. The $S^5$ is
described by coordinates $X^1\dots X^6$, and the rotation is in the
plane $X^5, X^6$. We write $Z=X^5+iX^6$, and let the other four
$X$ coordinates be called $X^i, i=1\dots 4$.  It was noted in
\cite{bmn} that (for appropriate choice of supergravity quantum)
the state in the dual CFT is created by the operator
$\mbox{tr}[Z^J]$, where $J$ is the angular momentum of the quantum.  It
was then argued that for large angular momentum $J$ we can
easily describe the excitations of the quantum that change it to a
stringy state. There are 8 bosonic excitations possible, and in a
light-cone gauge  4 are along the directions $X^i$ while 4 are
directions in the $AdS_5$.  Stringy excitations in the $X^i$
directions
are described in the dual theory by operators of the form
\be
\mbox{tr}[ ZZ\dots ZZX^{i_1}ZZ\dots ZZX^{i_2}ZZ\dots ZZ]
\label{qten}
\ee
Excitations in the $AdS_5$ directions are given by derivatives
$\partial_\mu, \mu=1\dots 4$.

Each of these insertions $X^i, \partial_\mu$ have $\Delta-J=1$. We do
not put insertions of $\bar Z$, which has $\Delta-J=2$.  While the
OPE of $X^i$ with $Z$ is regular in the free CFT, the OPE of $\bar Z$
with $Z$ gives
\be
\langle \bar Z(x_1) Z(x_2)\rangle\sim {1\over (x_1-x_2)^2}
\ee
The expectation is that the operators with $\bar Z$ insertions would
be renormalized to high dimensions when we go to the values of coupling
appropriate to the supergravity description, and so  these would
not be visible in the string vibration spectrum.

\subsection{String spectrum and energy scales for $AdS_3\times S^3\times T^4$}

Let us now consider the analogue of this problem for $AdS_3\times
S^3\times T^4$.  We take a supergravity quantum rotating
on the $S^3$ while staying at the origin in $AdS_3$. The spectrum of
string states was found in \cite{mets,bmn,tseytl}
\bea
\Delta&=&J+\sum_{m=-\infty}^{\infty}
\left\{{\tilde\omega}_m
a^{(+)\dagger}_m a^{(+)}_m+{\tilde\omega}_{-m}
a^{(-)\dagger}_m a^{(-)}_m+
{\tilde\omega}_m
{\tilde a}^{(+)\dagger}_m {\tilde a}^{(+)}_m+
{\tilde\omega}_{-m}
{\tilde a}^{(-)\dagger}_m {\tilde a}^{(-)}_m\right\}\nonumber\\
&+&\sum_{m=-\infty}^{\infty}
\sum_{i=1}^{4}
\left\{\left|\frac{L^2}{\alpha'}\frac{2m}{\Delta+J}\right|
c^{(i)\dagger}_m c^{(i)}_m\right\}+ferm
\label{MaldQuant}
\eea
Here $a$ give oscillators in the the $AdS$ directions,
$\tilde a$ give oscillations in the sphere directions,
and
$c_i$ give oscillations in the $T^4$. Further,
\be
{\tilde\omega}_m=\sqrt{\sin^2{\hat\alpha}+\left(\cos{\hat\alpha}+
\frac{L^2}{\alpha'}\frac{2m}{\Delta+J}\right)^2}\approx
1+\cos\hat \alpha({L^2 T_s\over 2\pi})({m\over J})
\label{qthree}
\ee
where  $L$ is a radius of the $AdS_3$, $T_S={1\over 2\pi \alpha'}$ is
the tension of the elementary string and the
approximation in the last step is for small mode numbers
$m\ll J$.

At leading order in $m/J$ we have the frequencies
${\tilde\omega}_m\approx 1$ for the oscillators in the $S^3$ and $AdS_3$
directions, which implies that these excitations have
\be
\Delta-J\approx 1
\label{wone}
\ee
The oscillators in the $T^4$ direction have ${\tilde\omega}\approx 0$ which
implies
\be
\Delta-J\approx 0
\label{wtwo}
\ee

The correction of order ${m\over J}$ in (\ref{qthree})  depends on
the value of $\hat \alpha$. But
the sigma model CFT at the
orbifold point is equally far in moduli
space from the D1--D5 background ($\hat\alpha=\pi/2$) and the NS1-NS5
background ($\hat\alpha=0$).
Any effects involving
$\hat
\alpha$ can only be obtained by studying  deformations of the
orbifold.  We will study the CFT only at the orbifold point, and it
is important to note that in such a study we should try to
reproduce only the values (\ref{wone}), (\ref{wtwo}) and not the
subleading corrections.

\subsection{CFT operators dual to string oscillators}

Consider a supergravity quantum rotating
on the $S^3$ while staying at the origin in $AdS_3$.  Letting this
quantum  be made from a suitable combination of $h_{ab}, B_{ab}$
($a,b$ are indices on $S^3$) we obtain a state with $h=j=\frac{n-1}{2},
~\bar h=\bar j=\frac{n-1}{2}$, which is a created in the dual CFT by a chiral
primary
\be
\sigma_n^{--}|0\rangle_{NS}
\ee
Let us write explicitly the permutation involved in the twist
operator $\sigma_n$.  We must finally sum over all choices of the
indices
involved in the permutation \cite{dvv,jevicki, lm1, lm2}, but to do this
we must first start by choosing definite values of the permuted
indices. Thus let the operator be
\be
\sigma_n^{--}=\sigma_{(12\dots n)}^{--}
\ee
where $(12\dots n)$ is the permutation  in the twist created by
$\sigma_n^{--}$.
Then we can write
\be
\sigma_{(12\dots n)}^{--}=\sigma^{--}_{(12)}\sigma^{--}_{(23)}\dots
\sigma^{--}_{({n-1},n)}
\label{threePrime}
\ee
So the chiral primary $\sigma_2^{--}$ which has a twist of order 2
acts like a building block for other chiral primaries, and is thus
similar to the operator $Z$ in the CFT dual to $AdS_5\times S^5$.

The operators $X_i, \bar Z$ in the $AdS_5\times S^5$ case were
$SO(6)$ rotates of the operator $Z$. In the
present case the corresponding rotation group is $SO(4)\approx
SU(2)_L\times SU(2)_R$.  There are two
vibrations on the sphere, which suggests that the string oscillators
in the sphere directions map to the CFT operators
\be
J_0^-\sigma_2^{--}, \bar J_0^-\sigma_2^{--}
\label{qone}
\ee
The analogue of $\bar Z$ is the operator with $SO(4)$ spin opposite
to the spin of $\sigma_2^{--}$, which is
\be
J_0^-\bar J_0^-\sigma_2^{--}
\label{qtwo}
\ee
Note that the operators (\ref{qone}) have $\Delta-J=1$ while the
operator (\ref{qtwo}) has $\Delta-J=2$; these are
    similar to the $\Delta-J$ values of the $X_i$ and $\bar Z$ respectively.

For the identification between string oscillators and CFT operators
to make sense we need to check in
addition that the OPE of the operators (\ref{qone}) with
$\sigma_2^{--}$ is regular, while the OPE of
(\ref{qtwo}) with
$\sigma_2^{--}$ is
singular.

In Appendix \ref{AppRev} we review the structure of twist operators.
Let us note some of the structure of chiral
primaries and their excitations. A twist
operator $\sigma_n$ just cyclically permutes $n$ copies of the $c=6$
CFT into each other. To make this into a
chiral primary with $h=j$ we must add charge, which can be done
by adding fractional modes of currents
$J^+_{-k/n}$.

Now consider the OPE of $J_0^-\sigma_2^{--}$ with $\sigma_2^{--}$.
Working independently in the
holomorphic and antiholomorphic sectors we find the apparent singularity
\be
[J^-_{0}\sigma_2^{--}](z)\sigma_2^{--}(w)\sim
\frac{1}{(z-w)^{1/3}}[J^-_{\frac{1}{3}}\sigma_3^{--}](w)
+\dots,
\label{qfive}
\ee
The operator  $[J^-_{\frac{1}{3}}\sigma_3^{--}](w)$ has $h-\bar h$
nonintegral. But recall that while fractional modes
are allowed for the left and right sectors, the allowed operators and
states in the CFT are only those with
$h-\bar h$ integral\footnote{For fermionic states $h-\bar h$ is
half integral.
$(h-j)-(\bar h-\bar j)$ is integral for all states.}
\cite{dasmathur, maldasuss}. Thus the singular
operator on the RHS of  (\ref{qfive}) does not exist in the CFT, and
the OPE is in fact regular
\be\label{OPEone}
[J^-_{0}\sigma_2^{--}](z)\sigma_2^{--}(w)\sim
[J_0^-\sigma_3^{--}](w)+\dots
\ee
Thus $J_0^-\sigma_2^{--}, \bar J_0^-\sigma_2^{--}$ are good
candidates for the `defects' that can be included
in the chain (\ref{threePrime}) to represent string oscillations:
\be
\dots\sigma^{--}_2\dots \sigma^{--}_2[J_0^-\sigma_2]\sigma^{--}_2\dots
\sigma^{--}_2[\bar
J_0^-\sigma^{--}_2]\sigma_2^{--}\dots
\label{qsix}
\ee

       Now consider the OPE of $J_0^-\bar J_0^-\sigma_2^{--}$
with
$\sigma_2^{--}$.  This time we get the leading singularity
\be\label{OPEtwoJ}
[{\bar J}^-_{0}J^-_{0}\sigma_2^{--}](z)\sigma_2^{--}(w)\sim
{1\over |z-w|^{2/3}}[J^-_{\frac{1}{3}}{\bar J}^-_{\frac{1}{3}}
\sigma_3^{--}](w)+\dots
\ee
The operator on the RHS has  $h-\bar h=0$ and is an allowed
operator in the theory.  Thus we should
not include $J_0^-\bar J_0^-\sigma_2^{--}$
in the chain (\ref{qsix}), which is analogous to not allowing $\bar
Z$ in the case of $AdS_5\times S^5$. The combination $Z\bar Z$ can be
contracted away to the identity; similarly
$(J_0^-\bar J_0^-\sigma_2^{--}) (\sigma_2^{--})$ can be contracted to
the identity and thus removed from the
chain.

In Appendix \ref{AppCFT} we discuss other OPEs that help support the picture
developed here.

\subsection{Mapping string oscillators to CFT operators}

The string vibrations in the sphere directions were mapped above to
$J^-_0\sigma_2^{--}, \bar
J^-_0\sigma_2^{--}$. Following the ideas of \cite{bmn} the
oscillators in the $AdS_3$ directions should be
mapped to $\partial_z, \partial_{\bar z}$, which upon acting on
$\sigma_2^{--}$ can be written as
\be
L_{-1}\sigma_2^{--}, \bar L_{-1}\sigma_2^{--}
\label{qel}
\ee
These operators also have $\Delta-J=1$.\footnote{See \cite{DasLeigh} for an
alternative interpretation of the
$AdS$ oscillators in $AdS_5\times S^5$.}

There are 4 fermionic vibrations that have have $\Delta-J=1$, and
following the above pattern it is natural to
identify these as
\be
G^{-,1}_{-{1\over 2}}\sigma_2^{--}, ~~G^{-,2}_{-{1\over
2}}\sigma_2^{--}, ~~\bar G^{-,1}_{-{1\over
2}}\sigma_2^{--},~~\bar G^{-,2}_{-{1\over 2}}\sigma_2^{--}
\label{qtw}
\ee

We can place the above `modified $\sigma_2$' operators (\ref{qone}),
(\ref{qel}), (\ref{qtw}) at various points
along the chain (\ref{threePrime}) and consider the linear
combinations of these possibilities that correspond
to definite Fourier harmonics along the chain \cite{bmn}. This gives
the map from string oscillators to CFT
operators:
\bea\label{AdSphIdent}
{\tilde a}^{(+)\dagger}_m|J\rangle&\rightarrow& \sum_{k=1}^J
e^{2\pi imk/J}(\sigma_2^{--})^{k-1}
[J^-_{0}\sigma_2^{--}](\sigma_2^{--})^{J-k-1}\nonumber\\
{\tilde a}^{(-)\dagger}_m|J\rangle&\rightarrow& \sum_{k=1}^J
e^{2\pi imk/J}(\sigma_2^{--})^{k-1}
[{\bar J}^-_{0}\sigma_2^{--}](\sigma_2^{--})^{J-k-1}\nonumber\\
b^{(+)i\dagger}_m|J\rangle&\rightarrow& \sum_{k=1}^J
e^{2\pi imk/J}(\sigma_2^{--})^{k-1}
[G^{-,i}_{-\frac{1}{2}}\sigma_2^{--}](\sigma_2^{--})^{J-k-1},\quad i=1,2
\nonumber\\
b^{(-)i\dagger}_m|J\rangle&\rightarrow& \sum_{k=1}^J
e^{2\pi imk/J}(\sigma_2^{--})^{k-1}
[{\bar G}^{-,i}_{-\frac{1}{2}}\sigma_2^{--}](\sigma_2^{--})^{J-k-1},\quad i=1,2
\nonumber\\
a^{(+)\dagger}_m|J\rangle&\rightarrow& \sum_{k=1}^J
e^{2\pi imk/J}(\sigma_2^{--})^{k-1}
[L_{-1}\sigma_2^{--}](\sigma_2^{--})^{J-k-1}\nonumber\\
a^{(-)\dagger}_m|J\rangle&\rightarrow& \sum_{k=1}^J
e^{2\pi imk/J}(\sigma_2^{--})^{k-1}
[{\bar L}_{-1}\sigma_2^{--}](\sigma_2^{--})^{J-k-1}
\eea

After constructing the above state of the twist operators we must
symmetrize by summing
over  permutations of the $N$ copies of the CFT.  The cyclic
permutations of the indices $(1\dots
n)$ involved in the above operators yield the same state, and so
these permutations must be
added with uniform weight. This forces the total `momentum' of the
Fourier modes $m$ to add
up to zero in the CFT, which agrees with the similar condition that
arises for the string oscillators
from the fact that the total momentum along the string world sheet
must be zero.

\subsection{CFT excitations as oscillations of a `Fermi fluid'}

We can obtain some understanding of the CFT states created by the
operators (\ref{AdSphIdent})
      by taking the specific case
      $M_4=T^4$ and writing the currents
$J^a$ in terms of the fermions of the CFT. We have 4 real fermions in the
holomorphic sector, and 4 real fermions in the
antiholomorphic sector. The holomorphic fermions are grouped into a
doublet of complex fermions
$(\psi^+, \psi^-)$ which forms a spinor of
$SU(2)_L$.  The currents are
\bea
J^+&=&\psi^+(\psi^-)^\dagger\nonumber\\
J^-&=&(\psi^+)^\dagger\psi^-\nonumber\\
J^3&=&\frac{1}{2}(\psi^+(\psi^+)^\dagger+(\psi^-)^\dagger\psi^-)
\eea

Consider for simplicity the chiral primary $\sigma_n^{--}$ for $n$
odd.  The charge that
must be added to the twist operator
$\sigma_n$ to make it a chiral primary is carried by fermions with
fractional moding:
\bea
\sigma_n^{--}&=&[\psi^{+}_{-{1\over 2n}}\psi^{+}_{-{3\over
2n}}\dots \psi^{+}_{-{1\over 2}+{1\over
n}}][(\psi^{-})^\dagger_{-{1\over 2n}}(\psi^{-})^\dagger_{-{3\over  2n}}\dots
(\psi^{-})^\dagger_{-{1\over 2}+{1\over n}} ]\nonumber\\
&&\times [\bar \psi^{+}_{-{1\over 2n}}\bar\psi^{+}_{-{3\over
2n}}\dots \bar\psi^{+}_{-{1\over
2}+{1\over n}}][(\bar\psi^{-})^\dagger_{-{1\over
2n}}(\bar\psi^{-})^\dagger_{-{3\over
2n}}\dots (\bar\psi^{-})^\dagger_{-{1\over
2}+{1\over n}} ]~\sigma_n
\label{qfour}
\eea
(Note that $\psi^+$ and $(\psi^-)^\dagger$ both carry the positive
$SU(2)_L$ spin.)

Now consider a state created by some Fourier mode $m$ of the
insertion $J_0^-\sigma_2$ in the manner
shown in  (\ref{AdSphIdent}). The Fourier function is approximately
constant over a large number of
$\sigma_2$ operators in the chain, since we seek to describe low
energy oscillations of the string by these
CFT operators. But if we sum over with uniform weight the states
obtained by insertion of
$J_0^-\sigma_2^{--}$ over a certain part of the chain of length $k-1$ then we
get
\be
\sum \sigma_2^{--}\dots [J_0^-\sigma_2^{--}]\dots \sigma_2^{--}~\sim~
J_0^-\sigma_k^{--}
\label{qfift}
\ee

The operator
$J_0^-$ acting on
$\sigma_n^{--}$ takes a fermion
$\psi^+$  to a fermion $\psi^-$, at the same level,
but does not change the number of fermions.  The operator in
(\ref{qfift}) rotates the spin of the fermions
from one direction of the $S^3$ to another direction over the segment
of the chain of length $\sim n$.  We
thus picture the modes (\ref{AdSphIdent}) as follows. The fermions
given in (\ref{qfour}) form a `Fermi sea'
over the long circle of length $2\pi n$ which results when the twist
$\sigma_n$ joins $n$ copies of the CFT into
one copy.  The Fourier modes  (\ref{AdSphIdent}) give low energy
oscillations of this Fermi sea by rotating the
sea in different directions in different regions of the long circle.
    This situation is reminiscent of  the $c=1$ matrix
model where low energy
excitations of spacetime were given by oscillations of a
nonrelativistic Fermi fluid \cite{dasjevicki}.

The operators $L_{-1}$ behave as $L_{-n}$ on the $c=6$ CFT on the
long circle as $L_{-n}$. Their effect can be
seen as the effect of a diffeomorphism in the CFT, which affects both
bosonic and fermionic variables.  In this
case it is more convenient to bosonise the  fermions, so that we have
6 effective bosons. The diffeomorphism
creates a `Bogolyubov transformation' on the bosonic vacuum. (The
bosons arising from the fermions have a
charge $e^{i\vec e \cdot\vec \phi}$ and this operator is also shifted
in the obvious way under the
diffeomorphism.)

\subsection{ $T^4$ vibrations and the orbifold point}

The string vibrations on the torus have $\Delta-J\approx 0$.
For small $\Delta-J$ we have the following bosonic and fermionic
modes in the CFT
\bea
\partial X^i_{-{k\over n}}, ~&&\psi^+_{-{1\over 2}-{k\over n}}, \quad
(\psi^+)^\dagger_{{1\over 2}-{k\over n}},\quad
(\psi^-)^\dagger_{-{1\over 2}-{k\over n}}, \quad
\psi^-_{{1\over 2}-{k\over n}}\nonumber\\
\bar \partial X^i_{-{k\over n}}, ~~&&\bar\psi^+_{-{1\over 2}-{k\over n}},
\quad (\bar\psi^+)^\dagger_{{1\over 2}-{k\over n}},\quad
(\bar\psi^-)^\dagger_{-{1\over 2}-{k\over n}},
\quad \bar\psi^-_{{1\over 2}-{k\over n}},\quad
\label{wthree}
\eea

The modes $(\psi^+)^\dagger_{{1\over 2}-{k\over n}}$,
$\psi^-_{{1\over 2}-{k\over n}}$ and their antiholomorphic counterparts act
as annihilation operators for $k<{n\over 2}$, and they become creation
operators for $k\ge {n\over 2}$.
(Note that since we have  restricted attention to $k\ll n$ we
will not reach this point in the excitation spectrum under
consideration.)

The energy levels implicit in the mode number $-{k\over n}$ in
(\ref{wthree}) are valid only for the CFT at the
orbifold point.  In \cite{orbCFT} it was argued that the orbifold point
has $n_5=1, n_1=N$.  Looking at the string
frequencies (\ref{MaldQuant}) we note that if we set $n_5=1$ and also
$\hat\alpha=0$, then $L=\sqrt{\alpha'}n_5$ and for the $T^4$
oscillators we get
\be
\tilde\omega_n=\frac{2m}{\Delta+J}\approx \frac{m}{J}
\ee
which agrees with the energy levels in (\ref{wthree}) in the CFT.
Similarly, the  string modes in the $AdS_3$ and
$S^3$ directions have energies
\be
{\tilde\omega}_n\approx 1+\frac{m}{J}
\ee

In the CFT at the orbifold point we find from the behavior of free
fermion energy
levels that
\be
\sum_{k=1}^n e^{2\pi imk/n}J^-_0\sigma^{--}_{(k,k+1)}~\rightarrow
~ J^-_{-{m\over n}}\sigma_n^{--}
\ee
More generally the Fourier modes in (\ref{AdSphIdent}) give operators
$J^-_{-{k\over n} }, G^{-,i}_{-{1\over 2}-{k\over n}},  L_{-1-{k\over n}}$
and their antihomomorphic partners, so that
we get an exact agreement of energy levels.
This agreement of energies cannot be taken too seriously though since
at the  point $n_5=1, n_1=N$ it is unlikely that the
string is well described by its free oscillation frequencies (\ref{qthree}).

In general it appears to be more useful to think of the CFT
excitations as generated by Fourier modes of local
insertions as in (\ref{AdSphIdent}), rather than as modes like
$J^-_{-{k\over n}}$ applied to the twist $\sigma_n^{--}$. As we
deform the CFT away from the orbifold point we expect the picture in
terms of collective vibrations to be a robust
one, since it just uses the fact that locally along the chain we can
apply a symmetry transformation to the spin
of  the chain. The exact energy of the state
generated by such oscillations can on the other hand
depend on the moduli of the CFT, and so we should use the orbifold
point to reproduce just the leading values
($\Delta-J\approx 1 , \Delta-J\approx 0$) for the various oscillators
of the string.\footnote{In \cite{japanese} the
energy levels of the CFT at the orbifold point were compared to the
string levels at $\hat\alpha=0$, and it was
noted that only those levels in the CFT were found in the string
spectrum which were multiples of $n_5$.}

\section{Discussion}

Let us collect together the above results (and results in \cite{lm4})
to get a picture of the
relation between rotation in $AdS_3\times S^3$ and twist operators in
the dual CFT.

The geometry $AdS_3\times S^3$ is dual to the NS vacuum
$|0\rangle_{NS}$ of the CFT. In
this state the CFT (at the orbifold point) has  $N$ copies of a $c=6$
CFT with each copy
living on a spatial circle on length $2\pi$.   The operator
$\sigma_n^{--}$ joins $n$ of
these copies to give a $c=6$ CFT living on a long circle of length
$2\pi n$, and adds charge
to give $h=j, \bar h=\bar j$.  The dual of the state
$\sigma_n^{--}|0\rangle_{NS}$ is
expected to be a massless particle  at the center of $AdS_3$ rotating
on the $S^3$ with
$SU(2)\times SU(2)$ angular momentum $(j, \bar j)$ \cite{exclusion}.

But in the CFT state $\sigma_n^{--}|0\rangle_{NS}$ we have circles of
length $2\pi$ and
also circles of length $2\pi n$. In \cite{lm4} dual geometries were
constructed for the
Ramond sector states which arise from spectral flows of general  chiral
primary states
$\sigma^{\pm\pm}_{n_1}\sigma^{\pm\pm}_{n_2}\dots
\sigma^{\pm\pm}_{n_k}|0\rangle_{NS}$; these states   in general have circles of
several different
lengths. Each metric had a singular curve with a
     shape depending on the set $\{ k_i\}$ and the choices  of signs $\pm$.

In our present study of the NS sector we have two cases where
simplifications occur:

\bigskip

(a)\quad We take the CFT state $(\sigma_n^{--})^{N/n}|0\rangle_{NS}$
where all circles are
of the same length.  Since we have a large number $N/n$ of particles
in the dual
geometry we have to consider their back--reaction,  and find the appropriate
new geometry.

\bigskip

(b)\quad We take the CFT state $\sigma_n^{--}|0\rangle_{NS}$ where
$n\ll N$. In this case
since we have just one light particle in the geometry we can ignore
its back--reaction and
keep the geometry $AdS_3\times S^3$. In this approximation we can  look at the
stringy excitations of this quantum,  following the ideas of \cite{bmn}.

\bigskip

We started by looking for the geometries required in (a).  For $n\gg 1$
the massless particles
are pointlike. We first considered the assumption that the metric will take the
Aichelburg---Sexl form (\ref{5DflatAS}) near the singular line of massless
particles --- this metric is
isotropic to leading order near the singularity.  It is interesting
that the solution
(\ref{NewASmetr})--(\ref{NewASfield}) with this limiting behavior can be
found in closed form, and
that the linear order
solution turns out to be exact.

But the solution (\ref{NewASmetr}) does not appear to have the right
properties to be the dual of the
CFT state  $(\sigma_n^{--})^{N/n}|0\rangle_{NS}$ . Following the
analysis of \cite{lm4} we
note that in the CFT all copies of the CFT live on circles of length
$2\pi n$, and so low
lying excitation energies should come in multiples of ${2\over n}$.
Further, since all
excitations in the CFT travel at the speed of light around the
spatial circle we expect that
any excitation in the dual geometry will be periodic with period
$2\pi n L$.  But in the
geometry (\ref{NewASmetr}) the scalar wave equation did not separate 
so that the energy levels had
no simple form; further the time for a
particle  to start from infinity and return to infinity was not order
$\sim 2\pi n L$.

We note that the space $AdS_3\times S^3$ is not invariant under
rotations in all 5 spatial
directions, and so it is possible that the lowest energy
configuration for a given angular
momentum might not be isotropic  even near the singular line of
particles.  In the R
sector study of \cite{lm4} it was found that the metrics dual to the
spectral flow of the
states
$(\sigma_n^{--})^{N/n}|0\rangle_{NS}$ had the simple form found in
\cite{bal, mm}: the
geometries are locally $AdS_3\times S^3$ and the singular curve is a
circle. We thus
consider the solution (\ref{NSsolut1})--(\ref{NSsolut2}) which is a simple
generalization of the solutions of \cite{bal,mm}, and has the quantum numbers
to be dual to the state
$(\sigma_n^{--})^{N/n}|0\rangle_{NS}$.\footnote{We have not proved
directly that this
geometry with its conical singularity is a solution of IIB string
theory. By contrast in
\cite{lm4} we did prove that all solutions (which also had singular
curves) were true
solutions in string theory: we started with a string having momentum
and winding,
found its geometry, and then dualized to obtain D1-D5 metrics with a
singular curve.}
The conical defect makes the solution non--isotropic around the
singular circle.

For the geometries (\ref{NSsolut1}) we find that the scalar wave equation
separates, and we found
the energy eigenvalues (\ref{OmegaK}).  For the simple case $l=0$
(particle with no angular
momentum) we found that the energy levels (\ref{OmKnoRot}) were multiples of
${2\over n}{1\over L}$.  This agreed exactly with the CFT expectation, where
the scalar considered
is described by one  left mover  and one right mover, and each has
energy ${k\over n}$.
We see here the phenomenon that was also seen in \cite{lm4}.  Low
energies and long
travel times arise in the CFT from {\it long circles}. In the dual
geometries we do not have
long circles but instead {\it deep throats} which create a large
redshift between infinity
and $r\sim 0$  so that we again get very low energy modes.

We now consider problem (b).  In \cite{bmn} stringy excitations in
$AdS_5\times S^5$
were described in the dual CFT by adding `defects' $X^i$ in a chain
of operators $Z$. We
have found a similar map in the case $AdS_3\times S^3$, where the
basic member of
the chain is the chiral primary $\sigma_2^{--}$ and the `defects' are
given by symmetry
operators like $J^-_0$ acting on  $\sigma_2^{--}$.

The excitation energies for a string with high $J$ are to leading
order $\omega\approx 1$
in units of the
$AdS$ scale  (in the $AdS_3$ and $S^3$ directions). The corrections
to these frequencies
depend on the tension of the string, and thus on the point in moduli
space that we choose
for the geometry. The orbifold point can describe at best only one
point in moduli space,
so we do not expect to reproduce these corrections by working at the
orbifold point.

The chiral primary $\sigma_n^{--}$ has a
large number of fermions that carry its charge, and these form a
`Fermi sea' in the $c=6$
CFT living on the long circle of length $2\pi k$.  The operator
$J^-_0$ rotates the spin
direction of these fermions, and thus Fourier modes of $J^-_0$ along
the chain generate
low energy oscillations of the Fermi sea.  When we deform the theory
away from the
orbifold point we expect that this concept of collective oscillations
will persist and yield the
stringy oscillations of the chiral primary.

We can also combine problems (a) and (b) to consider stringy
excitations of a quantum
moving in the background dual to the state
$(\sigma_n^{--})^{N/n}|0\rangle_{NS}$. We
can give the string a small center of mass energy which takes it away
from the singular
circle;  this lets its low energy excitations be those of a string in
$AdS_3\times S^3$
without the singularity. We see that the redshift between infinity
and $r\sim 0$ lowers
all string excitation energies by a factor $n$, when these energies
are measured from the
boundary of
$AdS_3$. In the dual CFT we start with circles of length
$2\pi n$ instead of $n$, and thus find energies that are also lower
by a factor $n$.

\section*{Acknowledgments}

We are grateful to Sergey Frolov, Camillo Imbimbo and Arkady Tseytlin for 
useful
discussions. This work was supported in part by  DOE grant DE-FG02-91ER40690.

\appendix
\section{Energy levels  for scalar excitations.}
\renewcommand{\theequation}{A.\arabic{equation}}
\setcounter{equation}{0}
\label{AppWave}

Here we solve the wave equation for a massless scalar.
We  look for the solution of the Klein--Gordon equation
\be
\Box\Phi=0
\ee
in the metric (\ref{NSsolut1}).

We write
\be
\Phi(t,r,\chi,\theta,\psi,\phi)=\exp(-i\omega t+ip\psi+iq\phi+i\la\chi)
H(r)\Theta(\theta),
\ee

Then we get the following equations for the two functions $H$ and
$\Theta$ (see \cite{lmTube} for
details):

\bea\label{ProperRad}
&&\frac{1}{r}\frac{d}{dr}\left(r(\frac{r^2}{L^2}+\gamma^2)\frac{dH}{dr}\right)+
\left\{\frac{(\omega-\beta p/L)^2}{\frac{r^2}{L^2}+\gamma^2}-
\frac{(\la+\beta q)^2}{r^2}\right\}H-\Lambda H=0\\
\label{ProperAngul}
&&\frac{1}{\sin 2\theta}\frac{d}{d\theta}\left(\sin 2\theta
\frac{d\Theta}{d\theta}\right)
-\left\{
\frac{q^2}{\sin^2\theta}+\frac{p^2}{\cos^2\theta}\right\}\Theta=-\Lambda\Theta,
\eea

The angular equation involves the usual Laplacian on $S^3$, so we get
$\Lambda=l(l+2)$ with $l=0,1,2,\dots$. Assuming that
$\la+\beta q\ge 0$, we get the solution of the
radial equation regular at $r=0$ \cite{lmTube}:
\bea\label{RadSolut}
H(x)&=&x^{(\la+\beta q)/2\gamma}(x+\gamma^2)^{(\omega L-\beta p)/2\gamma}
F(a,a+l+1;c;-\frac{x}{\gamma^2}),\\
{\rm where}~~~~a&=&-\frac{l}{2}+\frac{\la+\beta q+\omega L-\beta
p}{2\gamma},\qquad
c=1+\frac{\la+\beta q}{\gamma}.
\eea
For large  $z$  we have
\bea
&&F(a,a+m;c;z)\approx \frac{\Gamma(c)(-z)^{-a-m}}{\Gamma(a+m)\Gamma(c-a)}
\sum_{n=0}^\infty\frac{(a)_{n+m}(1-c+a)_{m+n}}{n!(m+n)!}z^{-n}(\log(-z)+h_n)
\nonumber\\
&&\qquad +\frac{\Gamma(c)(-z)^{-a}}{\Gamma(a+m)}
\sum_{n=0}^{m-1}\frac{(a)_{n}\Gamma(m-n)}{n!\Gamma(c-a-n)}z^{-n}
\eea
We will not need an explicit form of the coefficients $h_n$, they can be
found for example in \cite{batem}.
We then find that normalizability at infinity requires that the finite sum
disappears from the last expression, which happens if either one of the
conditions:
\be
c-a=-k \quad\mbox{or}\quad a=-l-k-1 \qquad\mbox{where}\quad k=0,1,2,\dots
\ee
is satisfied. This gives
\be
\frac{\omega L-\beta p}{2\gamma}=\pm
\left\{\frac{l}{2}+1+\frac{\la+\beta q}{2\gamma}+k\right\},
\ee
We thus get the  frequencies:
\be\label{OmegaKAp}
\omega_k=\frac{\beta p}{L}\pm\left\{\frac{2\gamma}{L}(k+1+\frac{l}{2})+
\frac{\la+\beta q}{L}\right\},\qquad
k=0,1,2,\dots
\ee

For the discrete set of frequencies with positive sign in (\ref{OmegaKAp}) we
can rewrite the
solutions of the wave equation in more convenient form:
\bea
H(x)&=&x^{c-1}(x+\gamma^2)^{k+l/2+1}
F(c+k,c+k+l+1;c;-\frac{x}{\gamma^2})\nonumber\\
&=&x^{c-1}(x+\gamma^2)^{-c-k-l/2}
({\gamma^2})^{c+2k+l+1}F(-k,-(k+l+1);c;-\frac{x}{\gamma^2})
\eea
The hypergeometric function in the above expression becomes a finite polynomial
of degree $k$ and $H(x)\sim x^{-1-l/2}$ as $x\rightarrow\infty$. For the
frequencies with negative sign in (\ref{OmegaKAp}) the hypergeometric function
in (\ref{RadSolut}) is also a finite polynomial.

If $\la+\beta q\le 0$, then instead of (\ref{RadSolut}) we get:
\bea
H(x)&=&x^{-(\la+\beta q)/2\gamma}(x+\gamma^2)^{(\omega L-\beta p)/2\gamma}
F(a',a'+l+1;c';-\frac{x}{\gamma^2}),\\
a'&=&-\frac{l}{2}-\frac{\la+\beta q}{2\gamma}+
\frac{\omega L-\beta p}{2\gamma},\qquad
c'=1-\frac{\la+\beta q}{\gamma}.
\eea
and we get the frequencies
\be
\omega_k=\frac{\beta p}{L}\pm\left\{\frac{2\gamma}{L}(k+1+\frac{l}{2})-
\frac{\la+\beta q}{L}\right\}\qquad k=0,1,2,\dots.
\ee
Combining this with (\ref{OmegaKAp}), we can write the spectrum for a general
case as
\be\label{ResulFreq}
\omega_k=\frac{\beta p}{L}\pm\left\{\frac{2\gamma}{L}(k+1+\frac{l}{2})+
\left|\frac{\la+\beta q}{L}\right|\right\}\qquad k=0,1,2,\dots.
\ee

\section{Chiral primaries in the orbifold CFT}
\renewcommand{\theequation}{B.\arabic{equation}}
\setcounter{equation}{0}
\label{AppRev}

We briefly summarize the nature  of chiral primaries in the orbifold
CFT and  introduce the notation that we use in this paper.
Details can be found in \cite{lm1, lm2}.

We consider the bound states of $n_1$ D1 branes and $n_5$ D5 branes
in IIB string theory. We
set
\be
N=n_1n_5
\ee
The D5 branes are wrapped on a 4-manifold $M$, and thus appear as
effective strings in the
remaining 6 spacetime dimensions. $M$ can be $T^4$ or K3. The  D1 branes
and the effective strings from the D5 branes extend along a common
spatial direction $x_5\equiv y$, and
                 $y$ is compactified on a circle of length $2\pi R$.  The low
energy dynamics of this system is a
{\cal N}=(4,4) supersymmetric 1+1 dimensional conformal field theory (CFT).
The CFT has an internal R-symmetry $SU(2)_L\times SU(2)_R\approx SO(4)$.
This
symmetry arises from the rotational symmetry of the brane
configuration in the noncompact
spatial directions $x_1$, $x_2$, $x_3$, $x_4$. The group $SU(2)_L$ is
carried by the left movers in the
CFT and the group $SU(2)_R$ is carried by the right movers.

Consider this CFT at the `orbifold point' \cite{orbCFT}. Then the CFT is a 1+1
dimensional sigma model where
the target space is the orbifold $M^N/S_N$, the symmetric product of
$N$ copies of the
4-manifold $M$.

The $M^N/S^N$ orbifold CFT and its states can be understood in the
following way.  We take $N$
copies of the supersymmetric $c=6$ CFT which arises from the sigma
model with target space
$M$. The vacuum of the theory is just the product of the vacuum in
each copy of the CFT.  In
the  orbifold theory we find twist operators  $\sigma_n$ \cite{dvv,jevicki}.
The copies
$1, 2,
\dots n$ of the CFT  permute  cyclically into each other
$1\rightarrow 2\rightarrow
\dots
\rightarrow n
\rightarrow 1$ as we circle the point of insertion of $\sigma_n$.
(The other copies are not
touched, and we ignore them for the moment.) In this given twist
sector there are operators
with various values of
$j_3$.

For odd $n$ we start with
$\sigma_n$, which is just a twist operator that permutes the copies of $M$
around its insertion point. The dimension of $\sigma_n$ is
\be
h_n=\bar h_n={c\over 24}(n-{1\over n})={6\over 24}(n-{1\over
n})={1\over 4}(n-{1\over n})
\label{eight}
\ee
To raise the charge of the operator with minimum increase in
dimension consider the application of $J^+_{-\frac{1}{n}}$. The charge goes
up by one unit, while the dimension increases by only $\frac{1}{n}$. The next
cheapest operator is $J^+_{-\frac{3}{n}}$, and continuing to apply
the cheapest possible operator at each stage we construct the
chiral operator in this twist sector with lowest dimension and charge.
We will call it $\sigma_n^{--}$ \cite{lm2}:
\be
\sigma^{--}_n\equiv J^+_{-\frac{n-2}{n}}J^+_{-\frac{n-4}{n}}\dots
J^+_{-\frac{1}{n}}{\bar J}^+_{-\frac{n-2}{n}}{\bar J}^+_{-\frac{n-4}{n}}\dots
{\bar J}^+_{-\frac{1}{n}}\sigma_n
\ee
and it has
\be\label{ChMinDim}
h= j_3={n-1\over 2}, ~~\bar h= \bar j_3={n-1\over 2}
\ee

In the case of even $n$, the lightest operator $\sigma_n$ in a given twisted
sector has dimension $h_n=\bar h_n={n\over 4}$
(see \cite{lm2} for details), and the chiral primary with lowest dimension is
\be
\sigma^{--}_n\equiv J^+_{-\frac{n-2}{n}}J^+_{-\frac{n-4}{n}}\dots
J^+_{0}{\bar J}^+_{-\frac{n-2}{n}}{\bar J}^+_{-\frac{n-4}{n}}\dots
{\bar J}^+_{0}\sigma_n
\ee
Its dimension and charge are given by (\ref{ChMinDim}).

Each copy of the CFT has the SU(2) currents $J^{(i)a}, \bar J^{(i)a}$,
where the index $i$ labels the
copies.  Define
\be
J^a=\sum_{i=1}^n J^{(i) a}, ~~~ \bar J^a=\sum_{i=1}^n \bar J^{(i) a}
\ee
Then we can make three additional chiral primaries from $\sigma^{--}$:
\bea
\sigma_n^{+-}&=&J_{-1}^+\sigma_n^{--}, ~~~~~~~h=j_3={n+1\over 2}, ~\bar
h=\bar j_3={n-1\over
2}\nonumber \\
\sigma_n^{-+}&=&\bar J_{-1}^+\sigma_n^{--}, ~~~~~~~h=j_3={n-1\over 2},
~\bar h=\bar j_3={n+1\over
2}\nonumber \\
\sigma_n^{++}&=&J_{-1}^+\bar J_{-1}^+\sigma_n^{--}, ~~~h=j_3={n+1\over 2},
~\bar h=\bar j_3={n+1\over
2}
\eea
The chiral primaries $\sigma_n^{--}, \sigma_n^{+-}, \sigma_n^{-+},
\sigma_n^{++}$ correspond respectively
to  the
$(0,0)$, $(2,0)$,  $(0,2)$, $(2,2)$ forms from the cohomology of $M$. Both
$T^4$ and K3 have one form
of each of these degrees.

The operator $\sigma_1^{--}$ is just the identity operator in one
copy of the $c=6$ CFT. Thus for
the complete CFT made from $N$ copies we can write the above chiral
operators as
\be
\sigma_n^{\pm\pm}[\sigma_1^{--}]^{N-n}
\ee
It is understood here that we must symmetrize the above expression
among all permutations
of the $N$ copies of the CFT; we will not explicitly mention this
symmetrization in what follows.

More generally we can make the chiral operators
\be
\prod_{i=1}^k ~[~\sigma_{n_i}^{s_i, \bar s_i}~]^{m_i},
~~~~~~\sum_{i=1}^k n_im_i=N
\ee
where $s_i, \bar s_i$ can be $+,-$. This
gives the complete set of chiral primaries that result if we restrict
ourselves to the above
mentioned cohomology of
$M$.

\section{Some computations in the CFT}
\renewcommand{\theequation}{C.\arabic{equation}}
\setcounter{equation}{0}
\label{AppCFT}

In \cite{bmn} strings in $AdS_5\times S^5$ were argued to be dual to
to traces like (\ref{qten}). We have
argued that strings in $AdS_3\times S^3\times T^4$ are described by
similar `chains' of $\sigma_2^{--}$
operators with sparsely placed `defects' (eqn. (\ref{qsix}));  these
defects correspond to vibrating the string in
the 2 transverse sphere directions and 2 transverse $AdS$ directions.

For this identification to be valid we
must ensure that there are no other `defects' that can be included in
the chain, since every type of defect
corresponds to a string oscillator in the gravity theory.
Note that we allow only those operators to appear as defects that
have a regular OPE with the basic member
of the chain
$\sigma_2^{--}$ . In computing this OPE we have to be careful that
only operators with $h-\bar h$
integral are allowed operators in the CFT, so we discard apparently
singular terms in the OPE which result in
operators with fractional  $h-\bar  h$.

We have allowed $J^-_0, G^{-,i}_{-{1\over 2}}, L_{-1}$ (and their
right moving analogs) to act on
$\sigma_2^{--}$ to produce the allowed defects. We start by noting
that higher modes of these operators
produce singular OPEs with $\sigma_2^{--}$ and so do not give allowed defects:
\bea
&&[J^-_{-n}\sigma_2^{--}](z)\sigma_2^{--}(w)=-\frac{1}{(w-z)^{n}}
\sigma_2^{--}(z)[J_0^-\sigma_2^{--}](w)+\dots,\qquad n>0,\\
&&[G^{-,i}_{-(\frac{1}{2}+n)}\sigma_2^{--}](z)\sigma_2^{--}(w)=
-\frac{1}{(w-z)^{n}}
\sigma_2^{--}(z)[G^{-,i}_{-\frac{1}{2}}\sigma_2^{--}](w)+\dots,\quad n>0,\\
&&[L_{-(n+1)}\sigma_2^{--}](z)\sigma_2^{--}(w)=
-\frac{1}{(w-z)^{n}}\d_w\sigma_3^{--}(w)+\dots,\qquad n>0
\eea
Note that these operators $J^-_{-n},  G^{-,i}_{-{1\over 2}+n},
L_{-(n+1)}$  (with $n>0$) all have $\Delta-J>1$.

The operators that give allowed defects have $\Delta-J=1$. We also
have some other operators with
$\Delta-J=1$ but which do not give regular OPEs with $\sigma_2^{--}$
and which therefore do not give new
allowed defects. An example is $J^3_{-1}$:
\be\label{OPEj3Ins}
[J^3_{-1}\sigma_2^{--}](z)\sigma_2^{--}(w)\sim
\frac{1}{z-w}\sigma_3^{--}(w)+\dots
\ee

Next we consider the issue of zero modes. In the string theory the
chiral primary is represented by a
particular member of the  multiplet of massless particles. By
application of fermionic zero modes
$d_0^{i\dagger}$ to the string we can change this particle to one of
the other massless particles.  To see these
zero modes in the dual CFT take the case $M_4=T^4$ for concreteness.
There are 4 fermionic modes in the
$c=6$ CFT with $\Delta=J=0$: ~$\psi^{+}_{-\frac{1}{2}}$,
${\bar\psi}^{+}_{-\frac{1}{2}}$,  $(\psi^{-}_{-\frac{1}{2}})^\dagger$,
$({\bar\psi}^{-}_{-\frac{1}{2}})^\dagger$ and we can insert these in a
chain of $\sigma_2^{--}$ operators. But
\be
(\sigma_2^{--})^k [\psi^{+}_{-\frac{1}{2}}
\sigma_2^{--}](\sigma_2^{--})^{J-k}
\sim \psi^{+}_{-\frac{1}{2}}\sigma_{J+1}^{--}
\ee
So these operators can be applied to the chain only in the zeroth
Fourier harmonic: they do not give
additional types of local defects but only give a global change to
the entire operator.  Thus the string zero
modes map to the CFT zero modes (we write only the bosonic operators)
\bea
|J\rangle&\rightarrow&\sigma_{J}^{--}\nonumber\\
\left\{d_0^{i\dagger}d_0^{j\dagger}|J\rangle\right\}&\rightarrow&
\left\{\sigma^{+-}_{J-1},\quad \sigma^{-+}_{J-1},\quad
\psi^{+\alpha}_{-\frac{1}{2}}{\bar\psi}^{+\beta}_{-\frac{1}{2}}
\sigma^{--}_{J-1}\right\}\nonumber\\
d_0^{1\dagger}d_0^{2\dagger}d_0^{3\dagger}d_0^{4\dagger}|J\rangle
&\rightarrow&\sigma^{++}_{J-2}
\eea
Taking supersymmetry descendents of the above set gives the other
massless particles of the gravity
multiplet.

      Finally we note that we also cannot get new `defects' by applying
the above zero mode operators to
the allowed defects. Note that $\psi^{+}_{-{1\over
2}}(\psi^{-}_{-{1\over 2}})^\dagger=J^+_{-1}$.  As an
example we can check that
$J^-_0J^{+}_{-1}\sigma^{--}_2$ has a singular OPE with $\sigma_2^{--}$
\be
J^-_0J^{+}_{-1}\sigma^{--}_2(z)\sigma^{--}_2(w)\sim
\frac{1}{z-w}J_0^-\sigma_3^{+-}(w)+\dots
\ee

\end{document}